\documentclass[12pt]{article}
\usepackage{amssymb,latexsym}
\usepackage[mathscr]{eucal}
\usepackage{amsmath,amssymb}
\usepackage{latexsym}
\usepackage{bbm}
\usepackage{bm}
\usepackage{cite}

\newcommand{\beq}{\begin{equation}}
\newcommand{\eeq}{\end{equation}}
\newcommand{\beqn}{\begin{eqnarray}}
\newcommand{\eeqn}{\end{eqnarray}}

\newcommand{\nn}{\nonumber}

\numberwithin{equation}{section}

\newcommand{\R}{\mathbbm{R}}
\newcommand{\Z}{\mathbbm{Z}}
\newcommand{\C}{\mathbbm{C}}
\newcommand{\eps}{\epsilon}

\newcommand{\be}{\begin{equation}}
\newcommand{\ee}{\end{equation}}
\newcommand{\ba}{\begin{eqnarray}}
\newcommand{\ea}{\end{eqnarray}}
\newcommand{\bdm}{\begin{displaymath}}
\newcommand{\edm}{\end{displaymath}}

\def\b{\beta}
\def\a{\alpha}

\newcommand{\ie}{{\it i.e.\ }}
\newcommand{\eg}{{\it e.g.\ }}

\DeclareMathAlphabet{\mathpzc}{OT1}{pzc}{m}{it}
\def\bea{\begin{eqnarray}}
\def\eea{\end{eqnarray}}
\def\beas{\begin{eqnarray*}}
\def\eeas{\end{eqnarray*}}
\def\sla{\raise.15ex\hbox{$/$}\kern-.57em}

\def\bea{\begin{eqnarray}}
\def\eea{\end{eqnarray}}

\def\sla{\raise.15ex\hbox{$/$}\kern-.57em}
\def\ie{{\it i.e.}~}
\def\eg{{\it e.g.}~}
\def\ap{{\alpha^\prime}}

\def\a{\alpha}
\def\b{\beta}

\def\cA{{\cal A}}
\def\cB{{\cal B}}
\def\cC{{\cal C}}

\def\cF{{\cal F}}
\def\cG{{\cal G}}
\def\cH{\C^+}
\def\cI{{\cal I}}

\def\cL{\Lambda}
\def\cM{{\cal M}}
\def\cN{{\cal N}}

\def\cT{{\cal T}}
\def\cU{{\cal U}}

\def\cX{{\cal X}}

\newcommand{\ft}[2]{{\textstyle\frac{#1}{#2}}}

\begin{document}
\begin{titlepage}
\begin{flushright}
{ROM2F/2007/21}\\
{CERN-TH-2007/257}\\
\end{flushright}
\begin{center}
{\large \sc  Unoriented D-brane Instantons \\ vs \\ Heterotic worldsheet Instantons}\\
\vspace{1.0cm}
{\bf Massimo Bianchi$^{1,2}$} and {\bf Jose F. Morales$^2$}\\
$^1${\sl CERN TH Division \\ CH - 1211 Geneva }\\
$^2${\sl Dipartimento di Fisica, Universit\'a di Roma ``Tor Vergata''\\
 I.N.F.N. Sezione di Roma II\\
Via della Ricerca Scientifica, 00133 Roma, Italy}\\
\end{center}
\vskip 2.0cm
\begin{center}
{\large \bf Abstract}
\end{center}

We  discuss Fermi interactions of four hyperini generated by
``stringy'' instantons in a Type~I~/~Heterotic  dual pair on
$T^4/\Z_2$.

\vfill
\end{titlepage}
\tableofcontents

\section{Introduction }

 In the last year or so, significant progress has been achieved in the
understanding of non-perturbative superpotentials generated by
unoriented D-brane instantons in Type I theories and alike \cite{
Ibanez:2006da, Blumenhagen:2006xt, Florea:2006si,Bianchi:2007fx, Cvetic:2007ku,
 Argurio:2007vqa,Bianchi:2007wy,Ibanez:2007rs,
 Antusch:2007jd,Blumenhagen:2007zk,
Aharony:2007pr,Blumenhagen:2007bn, Cvetic:2007qj, Billo:2007py,
Aharony:2007db, Camara:2007dy,Ibanez:2007tu,
GarciaEtxebarria:2007zv,Petersson:2007sc, Blumenhagen:2007ip,
Akerblom:2006hx, Akerblom:2007uc}. D-brane instantons are
euclidean D-branes entirely wrapped along the internal
compactification manifold and intersecting the D-brane gauge
theory at a point in spacetime. The instanton dynamics can be
efficiently described by effective gauge theories governing the
low energy interactions of open strings ending on the Dp-brane
instanton \cite{Billo:2006jm, Bianchi:2007ft}. The case when the
brane instanton is parallel to the brane hosting the gauge theory
along the internal space can be associated to YM instantons with
self-dual field strength. We refer to them as ``gauge instantons".
When the instanton and gauge theory branes wrap different cycles
in the internal manifold they are called ``exotic" or ``stringy"
since a clear field theory interpretation is still missing.

Only under very special conditions a  D-brane instanton can
generate  a superpotential in four dimensions. First, orbifold and
orientifold actions should be combined in such a way that out of
the original instanton fermionic zero modes only two, to be
identified as the  superspace fermionic coordinates
$\theta_\alpha$,  survive the projections.
 Second, the superpotential should transform in such a way as to balance the non-trivial
 transformation properties of the classical instanton action under all $U(1)$'s
 coming from D-branes intersecting the instanton.
  The resulting superpotential  violates the $U(1)$ symmetries under which the instanton
  is charged, leading to couplings in the low energy action that are otherwise
  forbidden at the perturbative level.
 In \cite{
Ibanez:2006da, Blumenhagen:2006xt, Florea:2006si, Cvetic:2007ku,
Blumenhagen:2007zk, Argurio:2007vqa, Bianchi:2007fx,
Aharony:2007pr,  Blumenhagen:2007bn, Cvetic:2007qj, Ibanez:2007rs,
Bianchi:2007wy, Antusch:2007jd, Billo:2007py, Aharony:2007db,
Camara:2007dy,Ibanez:2007tu,
GarciaEtxebarria:2007zv,Petersson:2007sc, Blumenhagen:2007ip,
Akerblom:2006hx, Akerblom:2007uc},  brane-worlds exposing both
gauge and exotic instanton generated superpotentials
 have been worked out.

   D-brane instantons are not new to the string literature. They showed up in
  the study of string dualities where brane and worldsheet instantons were shown
  to be related to each other via strong/weak coupling
  duality\footnote{For review see \eg
  \cite{Kiritsis:2007zz, Kiritsis:1997gu, Kiritsis:1997hj}}.
    The archetypical example is the computation of thresholds corrections to BPS saturated
 couplings, such as $F^4$ terms, in toroidal compactifications of  Type I / Heterotic string theories.
  The $F^4$ coupling is corrected by  ED1 branes in Type I theory and worldsheet instantons
  in the Heterotic string and the instanton sums have been shown to precisely match on the two sides
\cite{Kiritsis:1997hf, Bachas:1997mc, Kiritsis:2000zi}.  The
instanton dynamics was described in terms of the effective gauge
theory governing the interactions of ED1-D9 strings
\cite{Danielsson:1996es, Bachas:1997kn, Bachas:1997xn,
Gava:1998sv}. Unlike in the previously discussed exotic situation,
in this case brane instantons do not violate perturbative
symmetries of the theory and the $F^4$-couplings  are corrected
both at the perturbative and non-perturbative level.  Similar
results have been found for ${\cal R}^2$-terms in type I theory on
$T^6$, where the infinite tower of ED5--instanton corrections can
be rephrased in terms of worldsheet instantons in type IIA theory
on $K3\times T^2$  \cite{Hammou:1999in}.

It is natural to ask whether superpotentials generated at the non-perturbative
level by stringy instantons have a worldsheet instanton counterpart.
 Aim of this work is to consider the simplest setting where such a map can be built.
 We will exploit Type I /Heterotic duality to  rephrase exotic ED1-instanton
 corrections in terms of worldsheet instantons in the Heterotic string. We believe that
 a clear dictionary between the two will set the rules for exotic multi-instanton calculus
 on firmer grounds.

 We start by considering  brane / worldsheet instanton corrections in a
 ${\cal N}=(1,0)$ Heterotic / Type I
 dual pair in $D=6$ after compactification on $T^4/\Z_2$. The present analysis has been partly inspired by  the elegant test of the duality between
 Heterotic string on $T^4$ and Type IIA on $K3$ performed by Kiritsis, Obers and Pioline \cite{Kiritsis:2000zi}.
 The Type I vacuum is the
 $U(16)\times U(16)$  model constructed in \cite{Bianchi:1990tb}
 and rederived in \cite{Gimon:1996rq},
 see also \cite{Bianchi:1990yu, Bianchi:1991eu, Angelantonj:1996mw,Dabholkar:1996pc, Gimon:1996ay, Polchinski:1996ry, Berkooz:1996iz}.
 In the heterotic dual, the $SO(32)$ group is broken by the orbifold projection
to $U(16)$.  To achieve exact duality, the Type I D5-branes,
giving rise to the second factor in the Chan-Paton group, are to
be democratically distributed over the sixteen fixed points,
 breaking the D5-brane gauge group $U(16)$ down to $U(1)^{16}$.  The 16
 $U(1)$ photons then become massive by eating
twisted matter leading to a massless spectrum that matches
precisely the one of the Heterotic string \cite{Berkooz:1996iz}.
  We will consider quartic couplings involving hypermultiplets coming from the twisted sectors
  in the heterotic string. When the four hypers are located at
  different fixed points
 this  coupling is absent in perturbation theory but is generated by ED1 / worldsheet
  instantons wrapping two-cycles intersecting the four fixed points.

  It would be nice to extend our analysis to Heterotic / Type I dual pairs with $\cN =1$ supersymmetry
  in $D=4$ dimensions. In particular the $T^6/\Z_3$ orbifold of \cite{Angelantonj:1996uy},
  reconsidered from the D-brane instanton perspective in \cite{Bianchi:2007fx, Bianchi:2007wy},
   would be the first natural candidate to focus on,
   possibly after inclusion of Wilson lines \cite{Cvetic:2000st, Cvetic:2000aq}.
  The results for Type I / Heterotic dual pairs on
  freely acting orbifolds
  \cite{Camara:2007dy} go along this line.

 \section{Type I/Heterotic duals on $T^4/\Z_2$}

The first instance of a rather non-trivial yet workable Heterotic
/ Type I dual pair is the $U(16)$ model with $\cN = (1,0)$
supersymmetry in $D=6$, emerging from a $T^4/\Z_2$ orbifold
compactification. On the Type I side the model was originally
discussed among many others in \cite{Bianchi:1990tb} and was then
rediscovered in terms of D-branes and $\Omega$-planes in
\cite{Gimon:1996rq}. The combined effect of orbifold and
unoriented projections requires the presence of 32 D5-branes in
addition to the usual 32 D9-branes. At the maximally symmetric
point, where all the D5-branes are on top of an orientifold
O5-plane
 and no Wilson lines are turned  on the D9's, the gauge
group is $U(16)\times U(16)$.  With respect to many models of the
kind \cite{Bianchi:1990tb, Bianchi:1990yu, Bianchi:1991eu,
Angelantonj:1996mw, Dabholkar:1996pc, Gimon:1996ay,
Polchinski:1996ry} this Type I model is peculiar in that it has
only one tensor multiplet, whose scalar component determines the
volume of $T^4/\Z_2$, that can be related via Type I / Heterotic
duality to the unique tensor multiplet, containing the dilaton in
the heterotic side \footnote{Additional tensor multiplets can be
contributed by heterotic M5-branes, \ie pointlike $E(8)$
instantons, which do not admit a fullfledged worldsheet
description. Pointlike $SO(32)$ instantons support vector
multiplets very much like pointlike D5-branes in Type I, \ie
inside D9-branes.}.

  Besides its geometrical action, the $\Z_2$ orbifold acts on the
 32 heterotic fermions $(\lambda^u,\lambda^{\bar{u}})$, $u,\bar{u}=1,..16$
 (accounting for the world-sheet current algebra) as $(i^{16},(-i)^{16})$.
 This action effectively breaks $SO(32)$ to $U(16)$. The resulting spectrum
 is summarized in table \ref{theterotic}. We recall that a hyper H in a representation
 ${\bf R}$ consists of two half-hypers (each containing a complex boson and a chiral fermion)
 transforming in the representation
 ${\bf R}+{\bf  R^*}$ .
 \begin{table}[htdp]
\begin{center}
\begin{tabular}{|c|c|}
\hline
Sector &  $U(16)$ \ representations\\
\hline
Untwisted &  ${\bf 1_0}$ G +${\bf 1_0}$ T+${\bf 256_0}$  V+($4\times{\bf  1_0}$ +${\bf 120_2}$ +${\bf 120^*_{-2}}$) H  \\
Twisted     &   $16\times {\bf 16_{-3}}$ H\\
\hline
\end{tabular}
 \caption{\footnotesize Massless spectrum of heterotic string on $T^4/\Z_2$. G,T, V,H refers
to gravity, tensor, vector and hyper multiplets of ${\cal
N}=(1,0)$ $D=6$ supersymmetry.} \label{theterotic}
\end{center}
\end{table}%
Notice that the absence of twisted singlets, the would-be blowing
up modes,  implies that one cannot resolve the (hyperkahler)
orbifold singularity and reach a smooth K3 compactification
without at the same time breaking $U(16)$ at least to $SU(15)$ by
giving VEV to the charged twisted hypers \cite{Honecker:2006qz}.

  At the maximally symmetric point, the Type I spectrum looks rather
different ( see table \ref{ttypeI}).
   \begin{table}[htdp]
\begin{center}
\begin{tabular}{|c|c|}
\hline
Sector &  $U(16)\times U(16)$ \ representations\\
\hline
Untwisted closed&  $({\bf 1_0},{\bf 1_0})$ G +$({\bf 1_0},{\bf 1_0})$  T+$4\, ({\bf 1_0},{\bf 1_0})$   H  \\
\underline{Twisted  closed}   &   $16\, ({\bf 1_0},{\bf 1_0})$ H\\
99-open strings & $({\bf 1_0},{\bf 256_0})$  V+$({\bf 1_0},{\bf 120_2}+{\bf 120^*_{-2}})$ H\\
\underline{55-open strings} & $({\bf 256_0},{\bf 1_0})$  V+$({\bf 120_2}+{\bf 120^*_{-2}},{\bf 1_0})$ H\\
59-open strings & $({\bf 16^*_{-1}},{\bf 16_1})$  H \\
\hline
\end{tabular}
\label{ttypeI}
\caption{\footnotesize Massless spectrum of Type I on $T^4/\Z_2$ at the
$U(16)\times U(16)$ point.  G,T,V,H refers
to gravity, tensor, vector and hyper multiplets of ${\cal N}=1$ $D=6$ supersymmetry.
 Underlined fields are Higgsed away when D5-branes are democratically distributed
 among the 16 fixed points .}
\end{center}
\end{table}%
 In order to make contact with the heterotic description, one has
to start with the 16  D5-branes democratically
distributed among the 16 fixed points thus breaking $U(16)$ down
to $U(1)^{16}$. The necessary 240 = (256 -16) hypers are provided
by the open string hypers in the $({\bf 120_2}+{\bf
120^*_{-2}},{\bf 1_0}) $. It is important to notice that the two vacua belong to
 disconnected components of the type I moduli space since there is no
continuous deformation connecting the $U(16)$ and $U(1)^{16}$
D5-brane distributions\footnote{We thank I.~Antoniadis for drawing
our attention onto this subtle point and A.~Uranga for clarifying
it.}. The 16 $U(1)$'s are anomalous. The corresponding vector
multiplets get massive by eating as many `neutral' closed string
hypers in a supersymmetric fashion. The surviving massless
spectrum precisely matches the heterotic spectrum in table
\ref{theterotic} once the surviving (anomalous) $U(1)$ is
identified with $Q_H=Q_9 + 4 \sum_i Q_5^i$ in such a way that the
charge of the ${\bf 16}$ hypers is  $1 +4(-1) = -3$. The reducible
$U(1)$ anomaly is disposed of by means of the $D=6$ version of the
GS mechanism \cite{Sagnotti:1992qw, Ferrara:1996wv,
Ferrara:1997gh}.

 We will consider instanton generated chiral couplings in the ${\cal N}=1$ low energy effective action.
 The main working example is the four-hyperini Fermi
 interaction\footnote{The Fierz identity $(\gamma_\mu)^{ab}
 (\gamma^\mu)^{cd}= \eps^{abcd}$ in $D=6$ explains the
 nomenclature.}
   \bea
S_{4Fermi} &=& \int d^6x d^4\theta  \, W_{f_1f_2f_3f_4}(q) \prod_{i=1}^4 \,H_{16,f_i}\\
   &=&    \int d^6x\, W_{f_1f_2f_3f_4}(q) \,\epsilon^{a_1..a_4} \,
\zeta_{a_1,f_1}^{u_1}    \,\zeta_{a_2,f_2}^{\bar{u}_2}
\,\zeta_{a_3,f_3}^{u_3}\, \zeta_{a_4,f_4}^{\bar{u}_4}
\delta_{u_1\bar{u}_2}\delta_{u_3\bar{u}_4}+ \ldots \label{4hypers}
\nn \eea with $H_{16,f_i}$ the hypermultiplet superfield in the
${\bf 16}_{-3}+{\bf 16^*}_{+3}$  of $U(16)$ localized at the fixed
point $f_i$
 and $W_{f_1f_2f_3f_4}(q)$ encoding the dependence on the neutral moduli $q$, parametrizing K\"ahler and complex structure
 deformations of $T^4/\Z_2$. Indices $u,\bar{u}=1,...16$, $a=1,..4$ run over the fundamental
 and spinor representation of $U(16)$ and $SO(5,1)$ respectively.
Dots in (\ref{4hypers}) refer to other four-fermi interactions
obtained by permutations of the $u_i,\bar{u}_i$ superscripts of
$\zeta_{a_i,f_i}^{u_i}$ and supersymmetry related terms. The
four-fermi coupling in (\ref{4hypers}) can be thought of as the
supersymmetric partner of  a two-derivative four-scalar term
describing the hypermultiplet metric \cite{Becker:1995kb}.

If the four fixed points are chosen to be different from one another, the coupling (\ref{4hypers}) is absent
to any order in perturbation theory since twisted fields lying at different fixed points
do not interact perturbatively. Such a term can instead be generated
 via ED1-brane or worldsheet instantons connecting the four fixed points.
 The contributions will be exponentially suppressed with the area of the $S^2\sim T^2/\Z_2$
 cycle wrapped by the instanton.
 In the next sections  we will determine the moduli dependence $W_{f_1f_2f_3f_4}(q)$ of the four hyperini coupling
 from string  amplitudes in the Heterotic
 and Type I theories.

Before entering the details of the computation, let us comment on
which kind of corrections one should expect in the two
descriptions.
   Moduli coming from the untwisted sector in Type I / Heterotic theory on $T^4/\Z_2$
   span  the symmetric space $[SO(4,4)/SO(4)\times
SO(4)] \times SO(1,1)$. The $SO(1,1)$ factor accounts for the
scalar in the tensor multiplet. In $D=6$, Heterotic / Type I
duality amounts to the following field identifications
\cite{Angelantonj:1996uy} \be \phi_{_H} = \omega_{_I} \qquad ,
\qquad \phi_{_I} = \omega_{_H} \ee where $\phi$ and $\omega$
denote the dilaton and volume modulus respectively. Supersymmetry
implies that there is no neutral couplings between vectors and
hypers \cite{Andrianopoli:1996cm, Andrianopoli:1996vr}. The gauge
coupling can only depend (linearly) on the scalar in the unique
tensor multiplet $\phi_{_H} = \omega_{_I}$, while $\phi_{_I} =
\omega_{_H}$ belongs to a neutral hyper. For this reasons the
hypermultiplet geometry should be tree-level exact in the
heterotic description, but may receive worldsheet instanton
corrections. On the other hand,  in the Type I description it can
receive both perturbative and non-perturbative corrections.

  \section{Heterotic Amplitude}

  In this section we compute  the four-hyperini Fermi interaction
   in the heterotic description. The
  relevant heterotic string amplitude is
 \bea
 && \cA^{Het}_{4Fermi} (s,t) =
\int d^2z_3 \langle c \tilde{c} V^\zeta_{\bf 16} (z_1, \bar{z}_1)
c \tilde{c} V^\zeta_{\bf 16^*} (z_2, \bar{z}_2) V^\zeta_{\bf 16}
(z_3, \bar{z}_3) c \tilde{c} V^\zeta_{\bf 16^*}
(z_4, \bar{z}_4) \rangle\nn \\
&&= \zeta_{a_1,f_1}^{u_1} \zeta_{a_2,f_2}^{\bar u_2}
\zeta_{a_3,f_3}^{u_3} \zeta_{a_4,f_4}^{\bar{u}_4}
\epsilon^{a_1...a_4}\left[\delta_{u_1 \bar u_2}  \delta_{u_3
\bar{u}_4}{\cal A}^{(12|34)} _{f_i}+ \delta_{u_1 \bar u_4}
\delta_{u_3 \bar{u}_2}{\cal
A}^{(14|32)}_{f_i}\right]\label{stringamp} \eea Notice that ${\cal
A}^{(12|34)}_{f_i}$ and  ${\cal A}^{(14|32)}_{f_i}$ are related by
a simple relabelling of the fixed points $f_i$'s  therefore we can
restrict our attention onto the amplitude ${\cal
A}^{(12|34)}_{f_i}$ with color structure $\delta_{u_1 \bar u_2}
\delta_{u_3 \bar{u}_4}$.

The heterotic string vertex operators read
\bea
V^\zeta_{\bf 16} & =& \zeta^{\bar u}_a(p)\,e^{-\varphi/2}\, \sigma_f\, S^a  \tilde\Sigma_u
\, e^{ipX} \nn\\
 V^\zeta_{{\bf 16}^*} &=&
\zeta^u_a(p)\,e^{-\varphi/2}\, \sigma_f\, S^a  \tilde\Sigma_{\bar u}
\, e^{ipX}\label{hetver}
 \eea
 with
 \bea
\tilde\Sigma_{ u}= \prod_{v=1}^{16}  e^{  i(-{1\over 4}+\delta_{uv})\,\varphi_v}
 \quad\quad \tilde\Sigma_{\bar u}= \prod_{v=1}^{16}  e^{ i({1\over 4}-\delta_{uv})\,\varphi_v}
\eea
  The various fields entering the string vertices are defined as follows.
  $S^a$ are
  $SO(5,1)$
  spin fields,  $\varphi$ and $\varphi_u$ the bosonization
  of the superghost and $SO(32)$ gauge fermions respectively and $\sigma_f$  is
  the bosonic $\Z_2$-twist field.
  The contribution to the conformal dimension $(h,\bar{h})$ of the string vertex
operators from $S^a$,
  $ e^{-\varphi/2}$, $\tilde{\Sigma}_{u,\bar{u}}$ and $\sigma_f$ sum up
  to (1,1), {\it viz.}
  \be
  (h,\bar h)=(\ft38,0)+(\ft38,0)+(0,\ft34)+(\ft14,\ft14)=(1,1)
  \quad ,
  \ee
    as expected for a massless field. Finally $c,\tilde{c}$ are
   the ghost associated to the $SL(2,\C)$ invariance that allows to fix the positions $z_1,z_2,z_4$
of three vertices. The string amplitude will then depend on the
$SL(2,\C)$ invariant cross ratio $z$ defined as \be z={z_{12}
z_{34} \over  z_{13} z_{24}}
\ee
with $z_{ij}=z_i-z_j$.
    The worldsheet correlators  needed to compute the heterotic string amplitude read
   \bea
 && {\prod_i d^2 z_i\over dV_{SL(2,\C)}}=d^2z_3 \langle c\tilde{c}(z_1)
c\tilde{c}(z_2) c\tilde{c}(z_4)\rangle =d^2z_3  |z_{12}z_{14}z_{24}|^2= d^2z  |z_{13}z_{24}|^4\nn\\
&& \langle  \prod_{i=1}^4 \, e^{i k_i X}(z_i) \rangle =
 \prod_{i<j}  z_{ij}^{2\ap k_i k_j}
= |z|^{\ap s}\, |1-z|^{\ap t} \label{building}\\
&& \langle  \prod_{i=1}^4 \, S^{a_i} e^{-\varphi/2}(z_i) \rangle =
\epsilon^{a_1...a_4} \, \prod_{i<j}  z_{ij}^{-{1\over 2}} \nn\\
&&\langle \, \tilde\Sigma_{ u_1}(\bar{z}_1)\, \tilde\Sigma_{\bar u_2}(\bar{z}_2)\,
 \tilde\Sigma_{\bar u_3}(\bar z_3)\, \tilde\Sigma_{ \bar u_4}(\bar{z}_4)\rangle
= \prod_{i<j}  \bar z_{ij}^{-{1\over 2}}   \left(
\ft{1}{\bar z} \,\delta_{u_1 \bar{u}_2} \delta_{u_3 \bar u_4}+
\ft{1}{1-\bar z} \,\delta_{u_1 \bar{u}_4} \delta_{u_3 \bar u_2}
  \right)  \nn
   \eea
   Finally the four $\Z_2$-twist field correlator  is given by \cite{Dixon:1986qv}:
 \be
 \cC_{4\sigma}  = \langle \prod_{i=1}^4 \, \sigma_{f_i} (z_i,\bar{z}_i)
 \rangle =
 \cC_{qu} \,\cL_{cl} \left[^{\vec{f}_{12}}_{\vec{f}_{13}}\right]
\label{z4twist}
\ee
 with $\cC_{qu}$ and $\cL_{cl}$ the quantum and classical parts  of the correlator
 respectively.  The quantum part $\cC_{qu}$ is independent of the location of the
twist fields while the classical part $\cL_{cl}=\sum e^{-S_{\rm inst}}$ sums over
all possible wrapping of the string world-sheet passing through the 4 fixed points.

 We denote by $\vec{f}_i$ the locations of the four fixed points
 \be
 \vec{f}_i=
\ft12(\epsilon^1_i, \epsilon_i^2, \epsilon_i^3, \epsilon_i^4) \quad\quad \epsilon_i^a= 0,1
\ee
and by $\vec{f}_{ij}=\vec{f}_i-\vec{f}_j$ their relative positions.
 In order to get a non-trivial coupling the $\vec{f}_i$ should satisfy the
 selection rule
 \be
 \sum_{i=1}^4\, \vec{f}_i=\vec{0} \,~{\rm mod}\, 2
 \ee
The two pieces in (\ref{z4twist})
can be written in  terms of the Teichm\"uller
parameter $\tau(z)$ of the torus doubly covering the sphere with two $\Z_2$
branch cuts. The relation between the cross ratio $z$ and
$\tau(z)$ is coded in \be z = {\vartheta_3^4(\tau) \over \vartheta_4^4
(\tau) }
\ee
The quantum and classical parts of the correlator read \cite{Dixon:1986qv}
 \bea
 \cC_{qu} &=& 2^{-{8\over 3}}\, \prod_{i<j} |z_{ij}|^{-{1\over 3}}\,\tau_2^{-2} |\eta(\tau)|^{-8}
 \label{cclassqu}\\
 \cL_{cl}\left[^{\vec{f}_{12}}_{\vec{f}_{13}}\right] &=&
\sqrt{\det{G}} \sum_{{}^{\vec{n}\in {\Z} + \vec{f}}_{\vec{m}\in {\Z} +
\vec{h}}} e^{ - {\pi\over \tau_2} (\vec{n} \tau + \vec{m})^t G
(\vec{n} \bar\tau + \vec{m}) + {2} \pi i \vec{n} B \vec{m}}
 \nn
\eea
with $G$ and $B$ denoting the metric and
antisymmetric tensor on $T^4$.
Plugging (\ref{building}) and (\ref{cclassqu}) into (\ref{stringamp})
and using
\be
|z_{13} z_{24}|^4\prod_{i<j} |z_{ij}|^{-{4\over 3}}=|z(1-z)|^{-{4\over 3}}
\ee
one finds
 \bea
 {\cal A}^{(12|34)}_{f_i}(s,t)
 &=&\int {d^2z \over \tau_2^2 |\eta(\tau)|^8 } |z|^{\ap
s-{4\over 3}} |1-z|^{\ap t-{4\over 3}}
\,{1\over \bar z}\,
\cL_{cl} \left[^{\vec{f}_{12}}_{\vec{f}_{13}}\right]  \label{stringamp2}
\eea
Switching to the torus measure via
   \be
{ d z\over z} =  i \pi \vartheta_2^4(\tau) d\tau \quad\quad
\label{dztau} \ee
and sending $s,t\to 0$\footnote{Regularization of IR divergences is understood when some fixed
points coincide.} one finds

\be
 {\cal A}^{(12|34)}_{f_i}(0,0)
 = \int_{{\cal F}_2}  {d^2\tau \over \tau_2^2}
 \,{\bar \vartheta_4^4 \over \bar\vartheta_3^4 }  \,
 \cL_{cl} \left[^{\vec{f}_{12}}_{\vec{f}_{13}}\right]\label{z4f}
\ee
   The integral runs over  the fundamental domain ${\cal F}_2$ of the index six subgroup $\Gamma_2$
   of $\Gamma=SL(2,\Z)$, defined as the  group of modular transformations leaving invariant
   $\vartheta^4_{2},\vartheta^4_{3},\vartheta^4_{4}$.
   Interestingly, as can be seen from (\ref{z4f}), the  four hyperini coupling receives contributions
   only from ground states of the right moving supersymmetric sector of the string.
Indeed one can think of  (\ref{z4f}) as a ``1/2 BPS saturated"  partition function
      counting   $\Z_2$-invariant wrappings of the worldsheet
  instanton   along $T^4/\Z_2$ passing through the four fixed points $\vec{f}_i$
  and carrying the right $U(16)$ charges to couple to the ${\bf 16}$ hyperini.
      The fact that only BPS worldsheet instantons contribute to the four-hyperini amplitude is
  not surprising since only   states preserving half of the eight supercharges of the theory
  can correct a chiral term in the $D=6$ $\cN = (1,0)$ low energy effective action.

   \subsection{The integral}

 Integrals of modular invariant functions times unshifted lattice sums over
the fundamental domain of the modular group $SL(2, \Z)$ where first computed by Dixon, Kaplunovsky and Louis
\cite{Dixon:1990pc}. Their classic results were later on generalized to integrals of $\Gamma_2$-modular forms
 and shifted lattice sums \cite{Mayr:1993mq, Mayr:1995rx, Kiritsis:1998en, Lerche:1998gz, Lerche:1999hg}.

In this section we closely follow the strategy advocated in
\cite{Kiritsis:2000zi} to evaluate the integral
  \be
 {\cal I} \left[^{\vec{f}}_{\vec{h}}\right]
 = \int_{{\cal F}_2}  {d^2\tau \over \tau_2^2}
 \,\Phi(\bar \tau) \,
 \cL_{cl} \left[^{\vec{f}}_{\vec{h}}\right]
 \quad\quad \,\Phi(\bar \tau)={\bar \vartheta_4^4 \over \bar\vartheta_3^4 }  \, \label{z4f2}
\ee

We start by recalling some basic facts about the modular group $\Gamma = SL(2, \Z)$ and
its order 2 finite index subgroups. A more detailed discussion can be found in the Appendix.

$\Gamma$ acts on  the upper half-plane $\C^+=\{ \tau_1\in (-\infty,\infty), \tau_2 \in
(0,\infty)\}$ by
 projective transformations of the modular parameter $\tau$
 \be \tau \rightarrow {a \tau +
b \over c\tau +d} \quad {\rm with} \quad\quad \left\{ a,b,c,d \in \Z ;{\rm det} \left(
\begin{array}{cc}
  a& b  \\
 c& d   \\
\end{array}
 \right) =1\right\}\label{sl2z}
\ee
 The group $\Gamma$ is generated by the transformations
 \be
 S:   \tau \to -{1/ \tau} \quad , \quad T: \tau\to \tau+1
 \ee
  under which the shifted latticed sums transform according to
  \bea
   \cL_{cl} \left[^{\vec{f}}_{\vec{h}}\right] (-{1\over \tau})&=&
  \cL_{cl} \left[^{\vec{h}}_{\vec{f}}\right](\tau)\nn\\
  \cL_{cl} \left[^{\vec{f}}_{\vec{h}}\right]( \tau+1)&=&
  \cL_{cl}\left[^{\: \: \vec{f}}_{\vec{h}+\vec{f}}\right](\tau)
 \eea
 $\Gamma_2$ is the subgroup of $\Gamma$ leaving $\cL [_{\vec{h}}^{\vec f}]$
 invariant   for any $f_i,h_i=0,\ft12$.
In a similar way $\Gamma_2^+,\Gamma_2^-,\Gamma_2^0$  can be identified
  as the subgroups of $\Gamma$ leaving invariant the shifted lattice sums
    $\cL [^{\vec{f}}_{\vec 0}]$, $\cL [^{\vec{0}}_{\vec h}]$
    and $\cL [^{\vec{f}}_{\vec f}]$, respectively.
 The generators of the modular subgroups and their fundamental domains ${\cal F}^{0,\pm}=\C^+/\Gamma_2^{0,\pm}$
 can be found by restricting the parity of $a,b,c,d$
 entering in (\ref{sl2z}). We summarize the results in the following table
 \bea
\left.
\begin{array}{|l|l|l|l|l|}
\hline
 G & \cL_{\rm invariant} & \tiny{ \left(
\begin{array}{cc}
  a & b \\
    c & d  \\
\end{array}
 \right)  }  & {\rm generators} & {\cal F}_G=\C^+/G\\
 \hline
 \Gamma & \cL [_{\vec{0}}^{\vec 0}]    &  \tiny{ \left(
\begin{array}{cc}
  \Z & \Z \\
    \Z & \Z   \\
\end{array}
 \right)  } & S, T   & {\cal F} \\
 \hline
 \Gamma_2^{+} & \cL [^{\vec{f}}_{\vec 0}]    &  \tiny{ \left(
\begin{array}{cc}
  2\Z+1 & 2\Z \\
    \Z & 2\Z+1   \\
\end{array}
 \right)  } & T^2,STS & \{1,S,T\}{\cal F} \\
\hline
 \Gamma_2^{-} & \cL [_{\vec{h}}^{\vec 0}]    &  \tiny{ \left(
\begin{array}{cc}
  2\Z+1 & \Z \\
    2\Z & 2\Z+1   \\
\end{array}
 \right)  } & T,ST^2S  & \{1,S,ST\}{\cal F} \\
\hline
 \Gamma_2^{0} & \cL [_{\vec{f}}^{\vec f}]    &  \tiny{ \left(
\begin{array}{cc}
  2\Z+1 & 2\Z \\
    2\Z  & 2\Z+1   \\
\end{array}
 \right)  },
 \tiny{ \left(
\begin{array}{cc}
  2\Z & 2\Z+ 1 \\
    2\Z + 1 & 2\Z   \\
\end{array}
 \right)  }
   & S,T^2   & \{1,T,TS\}{\cal F} \\
\hline
 \Gamma_2 & \cL [^{\vec{f}}_{\vec h}]    &  \tiny{ \left(
\begin{array}{cc}
  2\Z+1 & 2\Z \\
    2\Z & 2\Z+1   \\
\end{array}
 \right)  } & T^2,ST^2S   & \{1,T,S,TS,ST,TST\}{\cal F} \\
\hline
\end{array}
\right.
\nn
 \eea

   Now we are ready to perform the integral  (\ref{z4f}).
  Following \cite{Kiritsis:2000zi},
write the instanton sum as\footnote{Here we use the shorthand notation
$\vec{m} \,A\, \vec{n}=m_i \,A_{ij} \,n_j $}  \be
\cL\left[^{\vec{f}}_{\vec{h}}\right](\tau, \bar{\tau}; G, B) =
\sqrt{\det{G}} \sum_{{}^{\vec{n}\in {\Z} + \vec{f}}_{\vec{m}\in
{\Z} + \vec{h}}} e^{ - {\pi\over \tau_2} (\vec{n} \tau +
\vec{m})G (\vec{n} \bar\tau + \vec{m}) + {2} \pi i \vec{n} B
\vec{m}} \ee and introduce the induced metric and antisymmetric
tensors \be \cG_{\a\b} = M_\a^i G_{ij} M^j_\b \quad , \quad
\cB_{\a\b} = M_\a^i B_{ij} M^j_\b \ee where $M^i_\a = (n^i, m^i)$.
More explicitly
\be
 \cG_{11} = \vec{n}\, G\, \vec{n}\quad , \quad
\cG_{22} = \vec{m}\, G\, \vec{m} \quad , \quad \cG_{12} = \vec{n}\,
G\, \vec{m} \quad , \quad \cB_{12} = \vec{n}\, B\, \vec{m} \ee Also
define the induced K\"ahler $\cT (M)$ and complex structures $\cU
(M)$ \bea \cT (M) &=& \left(\cB_{12} + i
\sqrt{\det\cG} \right)\nn\\
\cU (M)& =&
\cG_{11}^{-1} \left(\cG_{12} + i \sqrt{\det\cG}\right) \label{tmum}
\eea
 In such
a way that the instanton sum becomes \be
\cL\left[^{\vec{f}}_{\vec{h}}\right] = \sqrt{\det{G}}
\sum_{M(\vec{f},\vec{h})} e^{2\pi i \cT(M)} e^{- {\pi \cT_2(M)
\over \tau_2 \cU_2(M)} |\tau - \cU(M)|^2} \ee with the sum running
over \be M=(\vec{n},\vec{m})\in ({\Z}^4 + \vec{f}, {\Z}^4 +
\vec{h}) \ee
 The sum over $M$ can be rewritten as a sum of representatives of
  $\Gamma_2$-orbits and $\Gamma_2$-images of the fundamental domain.
  This exploits the fact that contributions coming from two different $M$'s related by
   \be
M\to M\cdot  \left(
\begin{array}{cc}
  a & b \\
    c & d   \\
\end{array}
 \right)  =   \left(
\begin{array}{cc}
 \vec{n} a+\vec{m} c & , ~ \vec{n} b+\vec{m} d \\
\end{array}
 \right)  \label{modm}
\ee
can be rewritten as integrals of the same representative on two domains related by the
  corresponding  modular transformations  on $\tau$.
The different orbits are classified by the sub-determinants of $M$
and are indicated as  {\it trivial} orbit for $M=0$, {\it
degenerate}
    orbits for ${\rm det} M_{ij}=m_i n_j-m_j n_i=0~\forall i,j$  and {\it non-degenerate}
    orbits otherwise.

Let us in turn consider the various contributions

\subsubsection{ Trivial Orbit} $M=0$, only present for $\vec{f}=\vec{h}=0$,
corresponds to the (regulated) contribution of the massless
particles exchange. Since
$\cL\left[^{\vec{0}}_{\vec{0}}\right]$ is modular invariant
one can replace $\Phi(\bar \tau)$ with its images under $\Gamma/\Gamma_2$. One finds
 \bea
 \cI_{triv}\left[^{\vec{0}}_{\vec{0}}\right] &=&\sqrt{\det{G}}  \int_{\cF_2}
{d^2\tau \over \tau_2^2} \Phi(\bar \tau) = \sqrt{\det{G}}  \int_{\cF} {d^2\tau \over
\tau_2^2} \sum_{s=1}^6 \Phi( \bar \tau_s)\nn\\
&=&3\sqrt{\det{G}}  \int_{\cF} {d^2\tau \over
\tau_2^2} =  \pi^2 \sqrt{\det{G}} \label{itrivial0}
\eea
with $\tau_s$ the image of $\tau$ under the modular transformations in the quotient $\Gamma/\Gamma_2$ so that
\be
    \tau_s =(1,S,T,TS,ST,TST)\tau
\ee
 The third equality in (\ref{itrivial0}) follows from the identity
 \be
 \sum_{s=1}^6 \Phi(\bar \tau_s)=3\label{phi3}
\ee
 that can be easily checked using Jacobi's {\it aequatio identica satis abstrusa}.

\subsubsection{ Degenerate Orbits}

{\bf Degenerate Orbits} $\det(M_{ij}) = n_i m_j - n_j m_i = 0$
$\forall i, j$ with $M\neq 0$.
They are present whenever $f_i h_j - f_j h_i = 0$ $\forall i, j$
\ie only when $\vec{f}=0$ or $\vec{h}=0$ or $\vec{f}=\vec{h}$.
Since each of these three `twist structures' is invariant under a
larger subgroup $\Gamma_{2}^{\pm,0}$ (rather than $\Gamma_2$) it is convenient
to exploit this extra symmetry.

In the case ${\vec f} =0$, $\Lambda_{cl} [^{\vec 0}_{\vec h}]$ is
invariant under $\Gamma^-_2$ and the integral can be written as
 \be
\cI_{deg}\left[^{\vec{0}}_{\vec{h}}\right] = \int_{\cF_2} {d^2\tau
\over \tau_2^2} \Phi(\bar \tau)
\Lambda_{deg}\left[^{\vec{0}}_{\vec{h}}\right] = \int_{\cF^-_2}
{d^2\tau \over \tau_2^2} [\Phi(\bar \tau)  + \Phi (\bar \tau +1)]
\Lambda_{deg}\left[^{\vec{0}}_{\vec{h}}\right]
 \ee
 The degeneracy condition implies that elements in this orbit can be written as
 \be
 \vec{n}=c\, \vec{\ell} \quad\quad \vec{m}= d \,\vec{\ell}     \quad\quad \vec{\ell}\in \Z^4+\vec{h} \quad
 c\in 2\Z \quad d\in 2\Z+1  \label{degm}
 \ee
  Comparing with (\ref{modm}) we see that $M=(\vec n,\vec m)$ in (\ref{degm}) correspond to $\Gamma_2^-$-images
 of the representative
 \be
 M^-_{deg}=(\vec{0},\vec{\ell})
 \ee
 Such representative is invariant under $T$ ($a=b=d=1$, $c=0$)
 and therefore the remaining modular transformations unfold the
fundamental domain $\cF^-_2$  to the strip
$\C^+/T=\{\tau_2>0, |\tau_1|<1/2\}$.

Since with the chosen representative
$\Lambda_{deg}\left[^{\vec{0}}_{\vec{h}}\right]$ is independent of
$\tau_1$, the $\tau_1$ integral can be performed immediately and
projects onto the constant $d_0=2$ in the expansion of
$\Phi(\bar\tau) + \Phi
(\bar \tau +1) = \sum_n d_n \bar q^n$.
   Integration over $\tau_2$ then yields
\be \cI_{deg}\left[^{\vec{0}}_{\vec{h}}\right] = 2\sqrt{\det{G}}
\int_0^\infty {d\tau_2\over \tau_2^2} \sum_{{\vec{m}\in {\Z} +
\vec{h}}} e^{ - {\pi \over \tau_2} \vec{m} G \vec{m}} = {2
\sqrt{\det{G}} \over \pi}\sum_{{\vec{m}\in {\Z} + \vec{h}}}
(\vec{m} G \vec{m})^{-1}\label{ideg0} \ee Although independent of
$B$, this contribution has no counterpart in the type I
description at lowest perturbative order (disk). Indeed, as we
will see in the next section, there is no room for connected disk
amplitudes with such $\delta_{u_1\bar{u}_2}\delta_{f_1
f_2}\delta_{u_3\bar{u}_4}\delta_{f_3 f_4}$ tensor structure.

In the second case, $\vec{h}=\vec{f}$, $\Lambda_{cl} [^{\vec
f}_{\vec f}]$ is invariant under $\Gamma^0_2$ that includes $S$ in
the quotient $\Gamma^0_2/\Gamma_2$. The relevant integrals are \be
 \cI_{deg}\left[^{\vec{f}}_{\vec{f}}\right] = \int_{\cF_2}
{d^2\tau \over \tau_2^2} \Phi(\bar \tau)
\cL_{deg}\left[^{\vec{f}}_{\vec{f}}\right] = \int_{\cF^0_2}
{d^2\tau \over \tau_2^2} [\Phi(\bar \tau)  + \Phi (-1/\bar \tau)]
\cL_{deg}\left[^{\vec{f}}_{\vec{f}}\right]  \ee
 Using the relation
 \be
\cL_{deg}\left[^{\vec{f}}_{\vec{f}}\right](\tau) =
\cL_{deg}\left[^{\vec{0}}_{\vec{f}}\right]( ST\tau)
\ee
and observing that $\cF_2^0 = TS \cF_2^-$ one can map the
problem back to the previous case
\bea
 \cI_{deg}\left[^{\vec{f}}_{\vec{f}}\right] &=& \int_{\cF^-_2}
{d^2\tau \over \tau_2^2} [
\Phi(TS\bar \tau)  + \Phi (STS\bar \tau)]
\cL_{deg}\left[^{\vec{0}}_{\vec{f}}\right] (\tau,\bar\tau ) \nn\\
&=& \sqrt{\det{G}}   \int_{\rm strip} {d^2\tau \over \tau_2^2} \left(
{\vartheta_3^4\over \vartheta_2^4}-{\vartheta_4^4\over
\vartheta_2^4}\right)(\bar \tau) \sum_{{\vec{m}\in {\Z} +
\vec{h}}} e^{ - {\pi \over \tau_2} \vec{m}
G \vec{m}}\nn\\
&=& {\sqrt{\det{G}} \over \pi}\sum_{{\vec{m}\in {\Z} + \vec{h}}} (\vec{m} G
\vec{m})^{-1}\label{ideg1}
 \eea
As we will momentarily see, this contribution has a clear
counterpart at disk level in the Type I dual description.

 In the third case, $\vec{h}=0$,
 $\cL_{deg}\left[^{\vec{f}}_{\vec{0}}\right]$ is invariant under
$\Gamma^+_2$ that includes $STS$ in the quotient
$\Gamma^+_2/\Gamma_2$. The relevant integrals are \be
\cI_{deg}\left[^{\vec{f}}_{\vec{0}}\right] = \int_{\cF_2} {d^2\tau
\over \tau_2^2} \Phi(\bar \tau)
\cL_{deg}\left[^{\vec{f}}_{\vec{0}}\right] = \int_{\cF^+_2}
{d^2\tau \over \tau_2^2} [\Phi(\bar \tau)  + \Phi (STS \,\bar
\tau)] \cL_{deg}\left[^{\vec{f}}_{\vec{0}}\right] (\tau,\bar
\tau)\ee Using the relation \be
\cL_{deg}\left[^{\vec{f}}_{\vec{0}}\right](\tau) =
\cL_{deg}\left[^{\vec{0}}_{\vec{f}}\right](S\, \tau) \ee
 and observing that $\cF_2^+ = S \cF_2^-$
 one can map the problem back to the previous case
 \bea
 \cI_{deg}\left[^{\vec{f}}_{\vec{0}}\right] &=& \int_{\cF^-_2}
{d^2\tau \over \tau_2^2} [
\Phi(S\bar \tau)  + \Phi (ST\bar \tau)]
\cL_{deg}\left[^{\vec{0}}_{\vec{f}}\right] (\tau,\bar\tau ) \nn\\
&=&   \sqrt{\det{G}} \int_{\rm strip} {d^2\tau \over \tau_2^2} \left(
{\vartheta_2^4\over \vartheta_3^4}-{\vartheta_2^4\over
\vartheta_4^4}\right)(\bar \tau) \sum_{{\vec{m}\in {\Z} +
\vec{h}}} e^{ - {\pi \over \tau_2} \vec{m}
G \vec{m}}\nn\\
&=& 0\label{ideg2}
 \eea
 The vanishing result for $\vec{f}_{13}=0, \vec{f}_{12}\neq 0$ is consistent with the fact that
 no perturbative type I correction  to the four-hyperini
Fermi interaction will be found with this tensor structure.

\subsubsection{ Non Degenerate Orbit}

For simplicity we take\footnote{A general $\vec{f}$ can be always put
into this form  using an $SL(4,\Z)$ transformation on the $T^4$-moduli}   $\vec{f}=(0,0,0,f)$ .
The representative for these orbits may be taken to be
\bea
  n^i= 0   &&  \quad\quad    i=1,...p\nn\\
  m^i<n^i &&  \quad\quad    i=p+1,..4
\eea
 for some $p$. They corresponds to the cases when $\det(M_{ij})= 0$ for $i,j\leq p$.
 Any other $M$ in these orbits can be reached by a $\Gamma_2$
transformation. Conversely, one can fix the representative and act
on $\tau$ with $\Gamma_2$ to enlarge the region of integration to the full
upper half plane $\C^+$. Notice that $\Gamma_2$ transformations  preserve
the integer/half-integer nature  of $(\vec{n},\vec{m})$.

The resulting integral is of the form \be
\cI_{ndeg}\left[^{\vec{f}}_{\vec{h}}\right] =  \sqrt{\det{G}}
\sum_{M } e^{2\pi i \cT(M)} 2^{4-1} \int_{\C^+} \Phi(\bar \tau)
e^{- {\pi \cT_2(M) \over \tau_2 \cU_2(M)} |\tau - \cU(M)|^2} \ee
where the factor $2^3$ accounts for the choices of signs of
$n^i,m^i$ for $i\neq \alpha$ and the sum runs over \bea
&& n^i=0     \quad\quad i=1,...p\nn\\
&& n^i\in \Z+f^i     \quad\quad i=p+1,..4\nn\\
&& m^i \in \Z+h^i    \quad\quad i=1..p\nn\\
&& m^i\in \Z~{\rm mod} ~n^i+h^i     \quad\quad i=p+1,...4
\eea

Expanding $\Phi(\bar \tau)$ in a power series \be \Phi(\bar \tau)
= \sum_\nu c_\nu \bar q^\nu \ee one can perform the gaussian
integration on $\tau_1$ that yields \bea
\cI_{ndeg}\left[^{\vec{f}}_{\vec{h}}\right] =
2^{3}\, \sqrt{\det{G}}\sum_{M } e^{2\pi i \cT(M)}  {\cU_2(M)^{1\over
2} \over \cT_2(M)^{1\over 2}} \sum_\nu c_\nu e^{-2\pi i \nu
\cU_1(M)} \times \nn
\\ \int_{0}^\infty {d\tau_2 \over \tau_2^{3/2}} e^{- {\pi \cT_2(M)
\over \tau_2 \cU_2(M)} [\tau -  \cU_2(M)]^2 } e^{ 2\pi \nu \tau_2
- \pi \tau_2 \nu^2 {\cU_2(M) \over \cT_2(M)}}  \eea

The integral over $\tau_2$ can be computed with the aid of \be
\int_0^\infty {dy \over y^{3/2}} e^{-a y - {b\over y}} =
\sqrt{\pi\over b} e^{-2\sqrt{ab}} \ee and yields \bea
\cI_{ndeg}\left[^{\vec{f}}_{\vec{h}}\right] &=& \sqrt{\det{G}}
\sum_{M } {2^{3}\over \cT_2(M)} e^{2\pi i \cT(M)}
\sum_\nu c_\nu e^{-2\pi i \nu \bar \cU(M)}  \nn\\
&=&  \sqrt{\det{G}} \sum_{M } {2^{3}\over \cT_2(M)} e^{2\pi i
\cT(M)} \Phi (\bar \cU(M))   \label{indegf} \eea Not surprisingly
the result is given by a sort of generalized Hecke operator acting
on the $\Gamma_2$-modular form $\Phi$ of weight zero.
  In the next section
we will argue that the same sum appears in the  ED1-instanton sum
in type I theory.

\subsubsection*{$f=h$ case}

 Formula (\ref{indegf}) describes the generic case when hyperini  come
  from four (in general different)  fixed points. However, a significant simplification
  takes place for particular choices of the fixed points.

  \begin{itemize}

  \item{$\vec{f}=\vec{h}=0$. All four hyperini located at the same fixed point.

With the aid of (\ref{phi3}) one finds
 \bea
 \cI_{ndeg} \left[^{\vec{0}}_{\vec{0}}\right] &=& \int_{\cF_2}
{d^2\tau \over \tau_2^2} \Phi(\bar \tau)
\Lambda_{ndeg}\left[^{\vec{0}}_{\vec{0}}\right] =  \int_{\cF}
{d^2\tau \over \tau_2^2} \sum_{s=1}^6
\Lambda_{ndeg}\left[^{\vec{0}}_{\vec{0}}\right] (\tau_s,\bar
\tau_s)
\, \Phi( \bar \tau_s)\nn\\
 &=& 3 \int_{\cF} {d^2\tau \over
\tau_2^2} \Lambda_{ndeg}\left[^{\vec{0}}_{\vec{0}}\right]  =3
\sum_{M } {2^{3}\, \sqrt{\det{G}} \over \cT_2(M)} e^{2\pi i \cT(M)}
\eea

  }

\item{$\vec{f}=\vec{h}\neq 0$. Hyperini in conjugate pairs located
at two different fixed points.

Now one finds
\bea
 \cI_{ndeg} \left[^{\vec{f}}_{\vec{f}}\right] &=& \int_{\cF_2}
{d^2\tau \over \tau_2^2} \Phi(\bar \tau) \Lambda_{ndeg}\left[^{\vec{f}}_{\vec{f}}\right]
=  \int_{\cF} {d^2\tau \over
\tau_2^2} \, \sum_{s=1}^6 \Lambda_{ndeg}\left[^{\vec{f}}_{\vec{f}}\right] \Phi ( \bar \tau_s)\nn\\
&=&  \int_{\cF} {d^2\tau \over
\tau_2^2} \left(\Lambda_{ndeg}\left[^{\vec{f}}_{\vec{f}}\right]+\Lambda_{ndeg}\left[^{\vec{f}}_{\vec{0}}\right]+\Lambda_{ndeg}\left[^{\vec{0}}_{\vec{f}}\right]\right)\nn\\
 &=& \int_{\cF} {d^2\tau \over
\tau_2^2} \left(\Lambda_{ndeg}\left[^{\vec{0}}_{\vec{0}}\right](G',B')-
\Lambda_{ndeg}\left[^{\vec{0}}_{\vec{0}}\right](G,B)\right)\nn\\
&=& 2^{3}\,
 \sqrt{\det{G}}\sum_{M } \left({e^{2\pi i \cT'(M)} \over
\cT_2'(M)} -{e^{2\pi i \cT(M)} \over \cT_2(M)}\right) \eea where
in the third line we have rewritten the $\Z_2$ shifted lattice sum
in terms of
 the unshifted lattice sum with rescaled metric and two form
   \be
   G'_{ij}=2^{-|h^{i}|  -|h^{j}| } G_{ij} \quad\quad B'_{ij}=2^{-|h^{i}|  -|h^{j}| } B_{ij}
   \ee

}

\end{itemize}

\section{Type I amplitude }

Unlike in the heterotic case, the Type I dilaton belongs to a
hypermultiplet and therefore the hypermultiplet geometry could
receive corrections both at the perturbative and non-perturbative
(ED-string) level. Yet some scattering amplitudes may vanish in
perturbation theory (at least in the low-energy limit) and only
receive contributions when non-perturbative effects are taken into
account. For instance \be \cA^{Type~I}_{4Fermi} = \langle
V^\zeta_{\bf 16}V^{\zeta}_{{\bf 16}^*} V^\zeta_{\bf 16}
V^{\zeta}_{{\bf 16}^*} \rangle \ee vanishes to all orders in
perturbation theory when the external states are all located at
different fixed points $f_i\neq f_j$ $\forall i,j$ because of
global $U(1)_{\rm D5}^{16}$ symmetry.
 When fixed points are
equal in pairs or even all equal (there are no other possibilities
for $\Z_2$) there is a perturbative contribution that should match
-- and indeed does so -- with the contribution of the `degenerate
orbit' in the heterotic description.
 In the following we will compute perturbative (disk) and ED1-instanton
 contributions to the four-hyperini Fermi interaction
 in the Type I description.

\subsection{The perturbative story}

In this section we compute the disk contribution in Type I theory.
 The disk with the four insertions of 59-string vertices has two D9 boundaries and two
(in principle different) D5 boundaries.
  The open string amplitude reads
  \bea
 &&\cA^{Type~I}_{4Fermi}  =
\int dx_3 \langle c  V^\zeta_{\bf 16} (x_1) c  V^\zeta_{\bf 16^*}
(x_2) V^\zeta_{\bf 16} (x_3) c  V^\zeta_{\bf 16^*}
(x_4) \rangle \label{stringampI}\\
&&= \zeta_{a_1,\bar{f}_1}^{u_1} \zeta_{a_2,f_2}^{\bar u_2}
\zeta_{a_3,\bar{f}_3}^{u_3} \zeta_{a_4,f_4}^{\bar{u}_4}\,
\epsilon^{a_1...a_4}\, \left[
\delta_{u_1\bar{u}_2} \delta_{f_2 \bar{f}_3} \delta_{u_3\bar{u}_4}
\delta_{f_4 \bar{f}_1}+\delta_{u_3\bar{u}_2} \delta_{f_2
\bar{f}_1} \delta_{u_1\bar{u}_4} \delta_{f_4 \bar{f}_3}
 \right]\,{\cal A}_{h}\nn
\eea The open string vertex operators are given by \bea
 V^\zeta_{\bf 16}  &=& \zeta^u_{a,\bar{f}}(p)\, S^a e^{-\varphi/2} \sigma_f e^{ipX}
\Lambda_{\bar{f}, u} \nn\\
 V^\zeta_{{\bf 16}^*} &=& \zeta^{\bar u}_{a,f}(p)\, S^a
e^{-\varphi/2} \sigma_f e^{ipX} \bar\Lambda_{\bar u,f} \eea Rather
than the heterotic fermions $\lambda$, the Type I vertex operators
involve Chan-Paton matrices $\Lambda_{u,f}$ in the bifundamental
of $U(16)_{\rm D9}\times U(1)^{16}_{\rm D5}$ yielding
 \be
\Lambda_{\bar{f}_1 u_1} \Lambda_{\bar{u}_2 f_2 }
\Lambda_{\bar{f}_3 u_3}\Lambda_{\bar{u}_4 f_4} =
\delta_{u_1\bar{u}_2} \delta_{f_2 \bar{f}_3} \delta_{u_3\bar{u}_4}
\delta_{f_4 \bar{f}_1}
\ee
 or
\be
\Lambda_{\bar{f}_1 u_1} \Lambda_{\bar{u}_4 f_4} \Lambda_{\bar{f}_3
u_3} \Lambda_{\bar{u}_2 f_2 } =\delta_{u_3\bar{u}_2} \delta_{f_2
\bar{f}_1} \delta_{u_1\bar{u}_4} \delta_{f_4 \bar{f}_3}
 \ee
 for the two possible cyclically
inequivalent orderings. They correspond to the two tensor
structures in (\ref{stringampI}).

 The integration runs over the real line with $x_1<x_2<x_3<x_4$.
  The correlators coincides with their counterpart in the
heterotic string but now insertions lie on the boundary of the
disk
    \bea
 && {\prod_i d x_i\over dV_{SL(2,\R)}}=dx_3\, \langle c(x_1)
c(x_2) c(x_4)\rangle =dx_3 \, (x_{12}x_{14}x_{24})^2 \nn\\
&& \langle  \prod_{i=1}^4 \, e^{i k_i X}(x_i) \rangle =
 \prod_{i<j}  x_{ij}^{2\ap k_i k_j}
= x^{\ap s}\, (1-x)^{\ap t} \nn\\
&& \langle  \prod_{i=1}^4 \, S^{a_i} e^{-\varphi/2}(x_i) \rangle =
\epsilon^{a_1...a_4} \, \prod_{i<j}  x_{ij}^{-{1\over 2}}\nn\\
&& \langle \prod_{i=1}^4 \, \sigma_{f_i} (x_i)
 \rangle =
 \cC_{qu}(x)  \,\cL_{cl} \left[^{\vec{0}}_{\vec{h}}\right](x)
 \label{buildingI}
  \eea
The amplitude is expressed in terms of the $SL(2,\R)$ invariant ratio
    \be
     x ={x_{12} x_{34} \over  x_{13} x_{24}} = {\vartheta_3^4(i t) \over \vartheta_4^4(it)}
     \ee
     with $t$ the modular parameter of the annulus doubly covering the disk.
    The classical and quantum part  of the four-twist correlator
    read
     \bea
 \cC_{qu} (x)&=& 2^{-{4\over 3}}\, \prod_{i<j} x_{ij}^{-{1\over 6}}\, t^{-2} \,\eta(i t)^{-4}
 \label{cclassquI}\\
 \cL_{cl} \left[^{\vec{0}}_{\vec{h}}\right] (x)& =& \sqrt{\det{G}} \sum_{\vec{m}\in \Z+h} e^{ -
{\pi\over  t}  \vec m
  \cdot G\cdot  \vec{m}}
 \nn
\eea The classical part counts windings of open strings connecting
D5-branes living at the fixed points 1 and 2. Plugging
(\ref{buildingI}) and (\ref{cclassquI}) into (\ref{stringampI})
one finds
 \bea
 {\cal A}_h (s,t)
 &=&\int {dz \over t^2 \eta(it)^4 } x^{\ap
s-{2\over 3}} (1-x)^{\ap t-{2\over 3}}\, \cL_{cl}
\left[^{\vec{0}}_{\vec{h}}\right] \label{stringamp2I} \eea
Switching to the annulus measure via
   \be
{ d x\over x} =  i \pi \vartheta_2^4(it ) dt \quad\quad
\label{dztau} \ee and sending $s,t\to 0$ one finds \bea
 {\cal A}_h (0,0)
 &=& \int_0^\infty  {dt \over t^2} \,
 \cL_{cl} \left[^{\vec{0}}_{\vec{h}}\right](it)= \sqrt{\det{G}}\int_0^\infty  {dt \over t^2} \,
   \sum_{{\vec{m}\in {\Z} + \vec{h}}} e^{ - {\pi \over t} \vec{m}
G \vec{m}} \nn\\
&=&
   {\sqrt{\det{G}}\over \pi}\sum_{{\vec{m}\in {\Z} + \vec{h}}} {1\over \vec{m}
G \vec{m}}
 \label{z4fI}
\eea Formula (\ref{z4fI}) perfectly matches  the dual heterotic
result (\ref{ideg1}) coming from worldsheet instantons in
degenerate orbits with tensor structure
$\delta_{u_1\bar{u}_2}\delta_{u_3\bar{u}_4}$ and fixed point
locations  $\vec f_{12}= \vec f_{13}$ ! It is quite remarkable
that for this particular configuration of fixed points,
 no other perturbative contributions beyond the disk
 are necessary to reproduce the degenerate orbits contributing to the
 heterotic amplitude. This should be a consequence of the 1/2
 BPS nature of the coupling. Let us notice, however, that the
heterotic contribution (\ref{ideg0}), though independent of $B$,
has no disk counterpart in the type I description. Indeed, there
is no room for connected disk amplitudes with the tensor structure
$\delta_{u_1\bar{u}_2}\delta_{f_1
f_2}\delta_{u_3\bar{u}_4}\delta_{f_3 f_4}$.

   \subsection{ED1-brane instantons}

  Finally we consider non-pertubative corrections to the four-hyperini coupling in the Type I description. The dynamics of ED1-instantons is described by the two-dimensional gauge theory governing the interactions of open strings ending on ED1-branes.
  Here we follow \cite{Gava:1998sv} where the gauge theory describing D1 bound states in
  toroidal compactifications of Type I theory was studied in great details
  and its dynamics was shown to match that of the fundamental Heterotic
  string (see also \cite{Danielsson:1996es, Bachas:1997kn, Bachas:1997xn, Bianchi:1998vq}).
  More precisely, the ED1 gauge theory was shown to flow in the infrared  to a symmetric product
  CFT that can be analyzed with standard orbifold techniques.
  This simpler IR description  was exploited to study the spectrum of BPS states of the theory.
    We start by reviewing the match for  $F^4$-terms  in type I/heterotic  theory  on $T^2$
     \cite{Bachas:1997mc} (here extended to $T^4$).
     Then we consider the closely relate BPS saturated four-hyperini terms in $T^4/\Z_2$.

  \subsubsection{$T^4$-case}

  The low energy dynamics of a bound state of $k$ D1-strings in ten-dimensional
  Type I theory is described by an $O(k)$ two-dimensional $\cN = (8,0)$
  supersymmetric gauge theory
  with the following matter content \cite{Danielsson:1996es, Bachas:1997kn}
  \bea
  X^I,S^a      &&  {\bf \ft12 k(k+1)} \nn\\
     S^{\dot a} &&   {\bf \ft12 k(k-1)} \nn\\
\lambda^{u}      &&    {\bf  k} \label{d1d9}
  \eea
   with $I=1,..{\bf 8_v}$, $a=1,..{\bf 8_s}$, $\dot a=1,..{\bf 8_c}$ and $u=1,...32$.
   $X^I,S^a$ fields come from D1-D1 open strings while $\lambda^u$ describe excitations of D1-D9 strings.
   The moduli space of the   theory, defined by the flatness condition $[X^I,X^J]=0$, is parameterized by diagonal
   matrices $X^I$ with components
    $X^I_t$, $t=1,...k$ (the D1 positions).
    In the infrared limit off-diagonal components of the adjoint matter $X^I$, $S^a$ and $S^{\dot a}$
    can pair up and become massive.
    This is not the case for the D1-D9 fermions $\lambda_t^u$
    that remain massless since D1's move always inside D9-branes. After integrating out the massive modes,
 one is left with the following field content
        \bea
  X_t^I,S_{t}^a , \lambda^u_{t}   \quad\quad t=1,...k  \label{cartan}
  \eea
   The Cartan fields in (\ref{cartan}) are defined up to
   transformations of the Weyl group
   $S_k\ltimes \Z_2^k$ of $O(k)$ with $S_k$ acting by permutations and $\Z_2$ acting
   by reflection on the $O(k)$-fundamentals $\lambda^u_{t} \to -\lambda^u_{t}$.
   The four possible $\Z_2$-holonomies of $\lambda_t^u$ along
the two cycles four spin structures of the $SO(32)$ fermions in
the heterotic string. The resulting field content is that of $k$
copies of the Green-Schwarz heterotic string moving in the target
space
   \be
   \cM:\quad\quad  (\R^6\times T^4)^k/S_k
   \ee
with light-cone coordinates chosen along a two-cycle inside $T^4$.
      The Type I / Heterotic dictionary can be further tested by considering $F^4$-couplings
   in the two descriptions. Insertions of $F^4$ can soak at most 8 fermionic zero modes
   and therefore this coupling will receive contributions only from 1/2 BPS excitations of the
   orbifold CFT. Notice that only even spin structures of $\lambda_t^u$ can contribute to this
   amplitude since $\lambda_t^u$ in  the odd spin structure contribute additional fermionic zero modes.
   This is in contrast with the recently found instanton generated superpotentials
   in orientifold braneworlds  \cite{
Ibanez:2006da, Blumenhagen:2006xt, Florea:2006si, Cvetic:2007ku,
Blumenhagen:2007zk, Argurio:2007vqa, Bianchi:2007fx,
Aharony:2007pr,  Blumenhagen:2007bn, Cvetic:2007qj, Ibanez:2007rs,
Bianchi:2007wy, Antusch:2007jd, Billo:2007py, Aharony:2007db,
Camara:2007dy, Ibanez:2007tu,
GarciaEtxebarria:2007zv,Petersson:2007sc, Blumenhagen:2007ip,
Akerblom:2006hx, Akerblom:2007uc}
   coming from $\lambda$-like fermionic zero modes appearing in the two-dimensional
   ED1-gauge theory only in the odd-spin structure.

   The string vertex operator for $F$ can be derived from
    the D1-D9 action and is given by
   \be
   V_F=F_{\mu\nu}^a \, {U_2\over k}\,\int d^2z \, \sum_{t=1}^k\, (X_t^\mu \partial X_t^\nu-S_t \gamma^{\mu\nu} S_t)(z)
   \lambda_t T^a \lambda_t\, (\bar z)
   \ee
   Notice that the right and left moving part of the vertex operator are given by the currents for
   $SO(8)$ transverse Lorentz group and
   $SO(32)$ gauge group. Choosing $F$ along the Cartan of these groups the string amplitude
   can be written in the simple form
   \be
   \langle V_F^4 \rangle ={U_2^4\over k^4}\,\partial^4_v \partial^4_{\bar v} \, \left.
   \langle e^{v\cdot J_{SO(8)}+\bar{v}\cdot J_{SO(32)} }\rangle\right|_{v=\bar{v}=0} =
  {U_2^4\over k^4}\,\partial^4_v \partial^4_{\bar v} \,  \left. \cX^{\rm D1}(v,\bar{v})  \right|_{v=\bar{v}=0} \label{ampvv}
   \ee
     with  $\cX(v,\bar{v})$ the weigthed partition function and $v,\bar{v}$ belonging to the Cartan
     of $SO(8)$ and $SO(32)$ respectively.

      In the following we evaluate $\cX^{\rm D1}(v,\bar v)$ in the symmetric product CFT.  We restrict ourselves   to   1/2-BPS contributions $\cX^{\rm D1}_{\rm BPS}(v,\bar v)\sim v^4 $
      since contributions from states preserving less supersymmetries (higher orders in $v$) will
      cancel from  (\ref{ampvv}).

   1/2-BPS states come from sectors in the symmetric product
   CFT with exactly 8 fermionic zero modes.
   We recall that twisted sectors of the symmetric product CFT are classified by the conjugacy classes
   of $S_k$
   \be
    [g]: \quad\quad (1)^{m_1} (2)^{m_2}..... (k)^{m_k} \label{gconj}
   \ee
    with $\sum_\ell  \ell \,m_\ell =k$ and $(\ell)$ referring to a cyclic permutation of length $\ell$.
    Each factor $(\ell)^{m_\ell}$ can be thought as $m_\ell$ copies of  ``short" strings of length $\ell$.
    States in this
    sector are projected by the centralizer
    \be
    {\cal C}= \prod_\ell  S_{m_\ell} \ltimes \Z_\ell^{m_\ell}
    \ee
      It is easy to check that the center of mass of any short string group $(\ell)^{m_\ell}$ in (\ref{gconj})
      is invariant under the orbifold projection and therefore leads to 8 fermionic zero modes.
      The only twisted sectors with exactly 8 fermionic zero modes are then
      those with a single $[g]=(\ell)^m$ factor with $m\ell = k$. This has to be projected by
      $(m) \times \Z_{\ell,{\rm diag}}^{s}$ with $s=0,...\ell -1 $. The weigthed BPS partition $\chi^{\rm D1}_{\rm BPS}(v,\bar v)$
      can then be evaluated by tracing over states in the twisted sectors labelled by $\ell,m,s$.

    We consider a D1-string wrapping on a $T^2$-cycle of $T^4$ specified by the
    two vectors $M_k=(\vec{k}_1,\vec{k}_2)$ each made out of four integers
    (with greatest common divisor 1). $\vec{k}_{1,2}$
    count the number of times the two 1-cycles of $T^2$ wind around the four
    1-cycles
    of $T^4$. The induced K\"ahler $\cT (M_k)$ and complex structures $\cU (M_k)$
    of this $T^2$-cycle are given by (\ref{tmum}).
    Now take $k=m \ell$ ED1-strings wrapping this two-cycle.
     Labelling
     fields in the symmetric  product CFT by $\Phi_{i,t_i}$ with $i=1,..m$, $t_i=1,..\ell$.
   The generic
   $\left[ (\ell)^m, (m)\times \Z_{\ell,{\rm diag}}^s\right]$-twisted boundary conditions  can be written
   as
      \bea
     && \Phi_{t_i,i}(\sigma_1+1,\sigma_2)= \Phi_{t_i+1,i}(\sigma_1,\sigma_2)\nn\\
      && \Phi_{t_i,i}(\sigma_1+U_1,\sigma_2+U_2)= \Phi_{t_i+s,i+1}(\sigma_1,\sigma_2)\label{bc}
      \eea
       Gluing the $k$-copies, one can form a single periodic field on a torus with
       K\"ahler structure $k \cT(M_k)$ and  complex structure
       $[m\, \cU(M_k)+s]/ \ell$.
       Using (\ref{tmum}) the results can be written in the suggestive form
       \bea
       \cU(\ell,m,s,M_k)&=&{m\, \cU(M_k)+s\over \ell}=\cU(M)\nn\\
        \cT(\ell,m,s,M_k)&=&\ell m \, \cT(M_k)=\cT(M)\label{umltml}
       \eea
    with
       \be
      M=M_k \cdot   \left(
\begin{array}{cc}
 \ell & s \\
    0 & m   \\
\end{array}
 \right)  \label{mmk}
       \ee
     Summarizing  the 1/2-BPS index $\cX^{\rm D1}_{\rm BPS}$
    for k D1-strings can be written in terms of the $k=1$ D1 ( i.e. the heterotic)
    partition function  $\cX_{\rm het}$ with modular parameter $\cU(M)$,
   weighted by the ED1-action $2\pi \cT(M)$.
   More precisely
      \be
      \cX^{\rm BPS}_{D1}(v,\bar v,T,U)=\sum_{k,M_k}\sum_{m\ell =k}\sum_{s=0}^{\ell-1}
      \, e^{2\pi i k T(M_k)}
      \cX_{\rm het}\left(m v, m \bar v ; {m\, \bar \cU(M_k)+s\over \ell}\right)\label{typeihet}
      \ee
       with
       \be
   \cX_{\rm het}\left(v,\bar v; \tau\right)={ \sqrt{\det{G}}\over \tau_2^4}\,{\theta_1(v)^4\over \eta^{12}}\,
   \sum_{\a=2,3,4}  {\bar \theta_\a^{16}(\bar v)\over \bar \eta^{24}}   \label{typeihet2}
      \ee
      Plugging (\ref{typeihet}) into (\ref{ampvv}) one finds
       \bea
   \langle V_F^4 \rangle &=& \sum_{k,M_k}
   \sum_{m\ell=k}\sum_{s=0}^{\ell-1}\, e^{2\pi i k \cT(M_k)}\,
      \cX^{\rm BPS}_{\rm het}\left({m\, \bar \cU(M_k)+s\over \ell}\right)\nn\\
      &=&\sum_{M} e^{2\pi i  \cT(M)}\,
    \cX^{\rm BPS}_{\rm het}\left(\bar \cU(M)\right)\
      \label{vf4f}
      \eea
         with
       \be
   \cX^{\rm BPS}_{\rm het}\left(\bar \tau\right)=  \partial^4_v \partial^4_{\bar v}\,  \cX_{\rm het}\left(v,\bar v,\bar \tau\right)\big|_{v=\bar v=0}
   =
   {\sqrt{\det{G}}\over \bar\eta^{24}}
   \left\{
   \begin{array}{c} \sum_{\a}\bar \theta_\a^{15}\bar \theta_\a^{''''} \\
    \sum_{\a}  {\bar \theta_\a^{14} (\bar\theta_\a^{''})^2}
   \end{array}
   \right.
    \label{chibps}
      \ee
      The two cases correspond to taking either all four $F$ equal or pairwise different
      elements of the $SO(32)$ Cartan subgroup.
      The result (\ref{vf4f}) precisely matches the contribution of non-degenerate orbits in the heterotic string computed in \cite{Bachas:1997mc} !

      \subsubsection{$T^4/\Z_2$-case}

   One can extend the type I / Heterotic dictionary described in the last section  to
    $T^4/\Z_2$.  The instanton dynamics is now governed by a gauge theory
    describing the excitations of unoriented strings connecting ED1, D5, and D9 branes.
    We focus again on the infrared dynamics where only the fields along the Cartan are relevant.

Let us describe the fermionic content of this theory (bosonic
fields follow from the residual $\cN = (4,0)$ worldsheet
        supersymmetry). Take a single $k=1$ ED-string wrapped on a two-cycle
        inside $T^4$.
        The $\Z_2$ projects out 4 of the 8 ED1-ED1 fermionic modes $S^a$.
        The surviving $a=1,..,4$ runs over the spinor
        representation of $SO(5,1)$.   In addition the ED1-ED9 modes $\lambda$'s transform
         with eigenvalues ``$i$". This can be
    seen by noticing that $\Z_2$ reflects the gauge vector (since the ED1 wraps $T^4/\Z_2$)
    that couples    in the ED1-D9 action linearly to $\lambda \lambda$.
    This is precisely the $\Z_2$-action in the $\lambda$ heterotic fermions and breaks
    $SO(32)$
    down to $U(16)$.
    Finally ED1-D5 open strings lead to one extra fermionic mode $\mu_{f_i}$ at each of the
    16 fixed points.  Indeed  ED1-D5 open strings have  8 ND directions and
    therefore the only massless excitation of this string comes from the fermionic Ramond ground state.

    A disk computation involving three twisted and one untwisted vertices
   shows that the hyperini $\zeta^u_{a,f_i}$ couple to all these fermionic
   modes via the four-fermi coupling
   \be
   {\cal L}_{4f}= \zeta^u_{a,f}  \mu_{ f }  \lambda_{u } S^a+ \zeta^{\bar u}_{a,f}
   \bar{\mu}_{f }  \lambda_{\bar u } S^a
   \ee
The four-hyperini coupling in the effective action can be computed by bringing
down four powers of
     ${\cal L}_{4f}$ in the gauge theory path integral.  Notice that this precisely soaks up the four
      fermionic zero modes $S^a$ leading to a non-trivial result.
      The resulting correlator is given by the insertion of  $V_\zeta= \mu_{ f }  \lambda_{u } S^a$
     in the gauge path integral.
This  vertex has the right
      $SO(5,1)\times U(16)_{\rm D9}$ quantum numbers
      to be identified with the heterotic twisted vertex (\ref{hetver})
      at fixed point $f$.   This suggests that the infrared dynamics
of the $k=1$ ED1-D5-D9 system can be described
      by the heterotic sigma model on $\R^6\times T^4/\Z_2$, trivially
generalizing the $T^4$
      result of the previous section.
Notice that unlike in the $T^4$ case the ED1 world-sheet wraps a
curved two-cycle $T^2/\Z_2$ inside K3. This makes the
ED1/heterotic dictionary less transparent and requires a
Green-Schwarz formulation of the heterotic theory with light-cone
coordinates chosen along the curved space-time two-cycle wrapped
by the instanton. Still, resorting to the covariant description,
(\ref{hetver}) is the operator in the CFT with the correct quantum
numbers supporting our proposal.
     This suggests that the infrared dynamics of the $k=1$ ED1-D5-D9 system can be described
     by the heterotic sigma model on $\R^6\times T^4/\Z_2$, trivially generalizing the $T^4$
     result of the previous section\footnote{More precisely,
     the transverse degrees of freedom live on $\R^6\times
     (T^2/\Z_2)_\perp$ while the light-cone coordinates are
     compactified on $(T^2/\Z_2)_\parallel$.}.

     For $k>1$ one finds $k$ copies of the heterotic string moving on the symmetric
     product target space   $(\R^6\times T^4/\Z_2)^k/S_k$.  We can now proceed
      like in the $T^4$-case.
       The two-cycle wrapped by the $k=1$ D-string
       inside $T^4$ is specified by
        $M_k=(\vec{k}_1,\vec{k_2})$  with $\vec{k}_{1,2}$ made out of
        integers with greatest common divisor 1. The induced Kahler and complex structure  of this two-cycle
        is $ \cT(M_k)$
        and
  $\cU(M_k)$ respectively.
        Using the fact
     that only twisted sectors with exactly four fermionic zero modes
     of $S^a$ contribute
     to the amplitude we restricted to $(\ell)^m$-twisted sectors projected by $\Z_{\ell,{\rm diag}}^s$.
       This folds the $k$ copies into
  a single field  on a worldsheet  with K\"ahler and complex structure
  $\cT(M)=k \,\cT(M_k)$ and
  $\cU(M)=[m\,\cU(M_k)+s]/\ell$.
  This is in perfect agreement with the heterotic result   (\ref{indegf}) for the four-hyperino
  coupling on $T^4/\Z_2$.
   Indeed using (\ref{mmk}) one can always rewrite the matrix $M$ describing the wrapping
  number of the heterotic string worldsheet in terms of a matrix $M_k$ specifying
  the two-cycle inside $T^4$ that the ED1 wraps and the integers
  $\ell,m,s$, with $m\ell = k$ and $s=0,...\ell -1$, specifying the twisted sector in the $k$-symmetric product orbifold CFT.

\section{Outlook}

In the present paper we exploited Heterotic / Type I duality on
$T^4/\Z_2$ to improve our understanding of stringy multi-instanton
calculus in theories with 8 supercharges. In particular we
 computed the four-hyperini Fermi interaction
 \be
 S_{4Fermi} =\int d^6 x d^4 \theta W_{f_1f_2f_3f_4} \prod_{i=1}^4 H_{16,f_i}
 \ee
 with $H_{16,f}$ the twisted hypermultiplet matter transforming in the ${\bf 16}_{-3},{\bf 16}^*_{+3}$ of
 $U(16)$. This amplitude  is tree-level exact at the perturbative level but it receives corrections
 from the infinite sum of 1/2 BPS worldsheet instantons connecting the fixed points $f_i$.

In the type I side, perturbative corrections are allowed only for
particular choices of the four fixed points where the hyperini are
located. In the cases where such perturbative corrections are
present at disk level we matched the result against the
contribution of degenerate orbits in the heterotic description.
When the four hyperini are located at different fixed points the
coupling is absent at any order in perturbation theory but
generated non-perturbatively via ED1-instantons wrapping two
cycles on $T^4/\Z_2$ connecting the four fixed points.
 This is the $D=6$ analog of the stringy generated superpotentials in ${\cal N}=1$
 theories recently studied in \cite{
Ibanez:2006da, Blumenhagen:2006xt, Florea:2006si, Cvetic:2007ku,
Blumenhagen:2007zk, Argurio:2007vqa, Bianchi:2007fx,
Aharony:2007pr,Blumenhagen:2007bn, Cvetic:2007qj, Ibanez:2007rs,
Bianchi:2007wy, Antusch:2007jd,Billo:2007py,
Aharony:2007db,Camara:2007dy,Ibanez:2007tu,
GarciaEtxebarria:2007zv,Petersson:2007sc, Blumenhagen:2007ip,
Akerblom:2006hx, Akerblom:2007uc}.

The rules for multi D-brane instanton counting in this setting
properly generalize those for toroidal compactifications. It is
beyond the scope of the present investigation to tackle the subtle
problem of the field theory interpretation of these stringy or
exotic instantons. A list of possible candidates are
Hyper-instantons \cite{Anselmi:1994bu}, Octonionic instantons
\cite{Fubini:1985jm}, Euclidean monopoles \cite{Witten:1994cg} or
IC-instantons \cite{Bonelli:1999it}.

 Finally, it is worth commenting on a somewhat unexplored mechanism for moduli
 stabilization for Type I and other unoriented theories
 like the one presently considered.
 It consists in the Higgsing of anomalous $U(1)$'s living on D-branes
 sitting at orbifold fixed points
 thanks to their  mixing with (twisted) RR axions.
 The efficiency of this mechanism in the present model is perfectly clear in the
 Heterotic description where all such states (vectors, axions
and their superpartners) are absent altogether from the massless
spectrum\footnote{M.~B. would like to thank E.~Kiritsis,
F.~Quevedo, B.~Schellekens and A.~Uranga for interesting
discussions and comments on this point.}. In $D=4$ a remnant of
the $D=6$ anomaly are massive (non-)anomalous $U(1)$'s
\cite{Antoniadis:2002cs, Anastasopoulos:2003aj } that may have
interesting generalized Chern-Simons couplings
\cite{Andrianopoli:2004sv, Anastasopoulos:2006cz, DeRydt:2007vg}.
It would be worth exploring further the effectiveness of
(non)anomalous $U(1)$ in removing axions from the massless
spectrum in a supersymmetric fashion and thus `stabilizing' the
corresponding moduli superfields. We could then start talking
about a {\it Petite Bouffe}!

\vskip 1.0cm
\begin{flushleft}
{\large \bf Acknowledgments}
\end{flushleft}
\vskip 0.5cm

\noindent It is a pleasure to thank P.~Anastasopoulos, G. Pradisi,
E. Kiritsis, F.~Fucito, for discussions. A preliminary version of
this work was presented by M.~B. at the {\it Workshop on recent
developments in string effective actions and D-instantons},
Max-Planck-Institut f\"ur Physik, Munich, November 14 -16, 2007.
M.~B. would like to thank the Organizers for their kind invitation
and for creating a very stimulating atmosphere. This work was
supported in part by the INTAS grant 03-516346, MIUR-COFIN
2003-023852, NATO PST.CLG.978785, the EU RTN grant
MRTN-CT-2004-512194.

\newpage

\section*{Appendix: Modular group and its subgroups}

In this appendix we collect some relevant information on the
modular group $\Gamma = SL(2,\Z)$ and its finite index subgroups
of order 2.

\subsection*{The modular group}

The modular group $\Gamma = SL(2,\Z)$ is a infinite discrete
group. $\Gamma = SL(2,\Z)$ acts on the upper half plane $\cH$
($\tau_2
>0$) by projective transformations \be \tau \rightarrow {a \tau +
b \over c\tau +d} \quad {\rm with} \quad a,b,c,d \in Z : ad-bc =1
\ee Under projective transformations \be \tau_2 \rightarrow
{\tau_2 \over |c\tau + d|^2}\ee

Its fundamental region $\cF = \cH/\Gamma$ is \be \cF = \{\tau :
\tau_2>0, |\tau_1|<1/2, |\tau|^2>1\} \ee $\Gamma = SL(2,\Z)$ is
generated by the two transformations $T$ and $S$ \be T \ : \ \tau
\rightarrow \tau +1 \quad , \quad S \ : \ \tau \rightarrow -1/\tau
\ee $T$ and $S$ satisfy \be S^2 = (-) 1 \quad , \quad (ST)^3 = 1
\ee

$\cF = \cH/\Gamma$ contains one cusp point $\tau = i \infty$ (a
`fixed point' of the parabolic element $T$)  and two elliptic
points $\tau=i$ (of order 2, fixed under $S$) and $\tau=\exp(2\pi
i/3) \approx \exp(\pi i/3)$ (of order 3, fixed under $ST$).

$SL(2,\Z)$ is the discrete version of the global conformal group
on the sphere $SL(2,C)$ or better its restriction to the disk
$SL(2,R)$. With this in mind it is important to observe that a
modular transformation with $c\neq 0$ can be conveniently written
in the form \be (c\tau' - a) = - {1\over c\tau + d} \ee
Transformations with $c=0$ are simply $T^\ell$ \ie rigid
translations. The above rewriting makes it clear that modular
transformations map circles centered at $-d/c$ on the $\tau_1$
axis into circles centered at $a/c$ on the $\tau'_1$ axis (which
actually coincides with the $\tau_1$ axis as a whole). For $c=0$
circles become vertical lines. In particular $T$ maps circle into
circle and vertical lines into vertical lines. $S$ maps circles of
radius $R$ centered in the origin into circles of radius $1/R$
centered in the origin. Moreover vertical lines $\tau_1 = r$ are
mapped by $S$ into circles of radius $1/2r$, centered at $\tau_1 =
-1/2r$ and thus passing through the origin, and {\it vice versa}.
These observations help following the maps of the (equivalent)
copies of the fundamental region $\cF$ (known as `hyperbolic
triangles', each containing a cusp and two elliptic points).

For our purposes it is necessary to introduce some finite index
subgroups of $\Gamma$: $\Gamma_2$, $\Gamma^+_2$, $\Gamma^-_2$,
$\Gamma^0_2$.

\subsection*{$\Gamma_2$ modular subgroup}
$\Gamma_2$ is the group of projective transformations with \be
\Gamma_2 \ : \ b,c = 0 ~{\rm mod} ~2\quad\quad  {\rm thus} ~a,d = 1 ~{\rm
mod} ~2 \ee $\Gamma_2$ is generated by $T^2$ and $ST^2S$.
$\Gamma_2$ is the subgroup of $\Gamma$ preserving all half-integer
spin structures. Its fundamental region has genus 0 and is
obtained from $\cF = \cH/\Gamma$ by the action of the elements of
the coset $\Gamma/\Gamma_2$ i.e. \be \cF_2 = \cH/\Gamma_2 = \{
1,T, S, TS, ST, TST\} \cF \ee A convenient representation for
$\cF_2$ is the region bounded by the vertical lines $\tau_1 = \pm
1$ and by the two circles $(\tau_1 \mp 1/2)^2 + \tau_2^2 = 1/4$.

This in particular means that $\Gamma_2$ is of index 6 in $\Gamma$
and contains 3 cusps at $\tau =0,1,i\infty$ and no fixed points.

\subsection*{$\Gamma^-_2$ modular subgroup}

$\Gamma^-_2$ is the group of projective transformations with \be
\Gamma_2 \ : \quad c = 0 ~{\rm mod}~ 2 \quad ~ {\rm thus} ~ a,d =
1 ~{\rm mod} ~2 \ee $\Gamma^-_2$ is generated by $T$ and $ST^2S$.
$\Gamma^-_2$ is the subgroup of $\Gamma$ preserving the
half-integer spin structure of $\vartheta_2$  i.e.
$[^{\tiny 0}_{\tiny h}]$. Its fundamental region has genus 0 and
is obtained from $\cF = \cH/\Gamma$ by the action of the elements
of the coset $\Gamma/\Gamma^-_2$ i.e. \be \cF^-_2 = \cH/\Gamma^-_2
= \{ 1, S, STS \} \cF \ee A convenient representation for
$\cF^-_2$ is the region bounded by the vertical lines $\tau_1 =
\pm 1/2$ and by the two circles $(\tau_1 \mp 1/2)^2 + \tau_2^2 =
1/4$. This in particular means that $\Gamma^-_2$ is of index 3 in
$\Gamma$ and contains 2 cusps at $\tau =0,i\infty$  and one fixed
points of order 2 at $\tau = (1+i)/2$.

\subsection*{$\Gamma^+_2$ modular subgroup}

$\Gamma^+_2$ is the group of projective transformations with \be
\Gamma_2 \ : \quad b = 0 ~{\rm mod} ~2~ \quad~ {\rm thus} \quad  a,d =
1 ~{\rm mod} ~2~ \ee $\Gamma^+_2$ is generated by $STS$ and
$ST^2S$. $\Gamma^+_2$ is the subgroup of $\Gamma$ preserving the
half-integer spin structure of $\vartheta_4$  i.e.
$[^{\tiny f}_{\tiny 0}]$. Its fundamental region has genus 0 and
is obtained from $\cF = \cH/\Gamma$ by the action of the elements
of the coset $\Gamma/\Gamma^+_2$ i.e. \be \cF^+_2 = \cH/\Gamma^+_2
= \{ 1, S, T \} \cF \ee A convenient representation for $\cF^+_2$
is the region bounded by the vertical lines $\tau_1 = \pm 1$ and
by the two circles $(\tau_1 \mp 1)^2 + \tau_2^2 = 1$.

This in particular means that $\Gamma^+_2$ is of index 3 in
$\Gamma$ and contains 2 cusps at $\tau =0,i\infty$ and one fixed
points of order 2 at $\tau = 1+i$.

\subsection*{$\Gamma^0_2$ modular subgroup}

$\Gamma^0_2$ is the group of projective transformations with \bea
\Gamma^0_2 \ : &&
 b+c = 0 ~{\rm mod}~ 2 \quad  {\rm and} ~ a+b = 1 ~{\rm mod} ~2 \eea
$\Gamma^0_2$ is generated by $S$ and $ST^2S$. $\Gamma^0_2$ is the
subgroup of $\Gamma$ preserving the half-integer spin
structure of $\vartheta_3$  i.e. $[^{\tiny f}_{\tiny f}]$.

A convenient representation for $\cF^0_2$ is the region bounded by
the vertical lines $\tau_1 = \pm 1$ and by the unit circle
$(\tau_1)^2 + \tau_2^2 = 1$.

Its fundamental region has genus 0 and is obtained from $\cF =
\cH/\Gamma$ by the action of the elements of the coset
$\Gamma/\Gamma^0_2$ i.e. \be \cF^0_2 = \cH/\Gamma^0_2 = \{ 1, T,
TS \} \cF \ee This in particular means that $\Gamma^0_2$ is of
index 3 in $\Gamma$ and contains 2 cusps at $\tau = \pm 1,
i\infty$ and one fixed points of order 2 at  ($\tau =i$ under
$S$).

It proves crucial for our manipulations to observe that the above
fundamental regions are related to one another: \be \cF_2^- = S
\cF_2^+ \quad , \quad \cF_2^+ = T \cF_2^0 \quad , \quad \cF_2^+ =
TS \cF_2^- \ee where $T^2 = 1$ (true for weak modular forms of
$\Gamma_2^\alpha$ with $\alpha =\pm, 0$ or nothing) as well as
$S^2=1$ has been used.


\providecommand{\href}[2]{#2}\begingroup\raggedright\endgroup

 \end{document}